\documentclass[aps,prl,reprint,superscriptaddress,nofootinbib,floatfix]{revtex4-2}

\usepackage{amsmath,amssymb,bm}
\usepackage{graphicx}
\usepackage{booktabs}
\usepackage{xcolor}
\usepackage{microtype}
\usepackage{setspace}
\usepackage[colorlinks=true,linkcolor=blue,citecolor=blue,urlcolor=blue]{hyperref}

\newcommand{\tr}{\mathrm{Tr}}
\newcommand{\Dtr}{D_{\mathrm{tr}}}
\newcommand{\rhoSS}{\rho_{\mathrm{ss}}}
\newcommand{\rhoTh}{\rho_{\mathrm{th}}}
\newcommand{\Liouv}{\mathcal{L}}
\newcommand{\Diss}{\mathcal{D}}
\newcommand{\ket}[1]{|#1\rangle}
\newcommand{\bra}[1]{\langle #1|}
\newcommand{\Tcal}{\mathcal{T}}

\begin{document}
\title{Strong Quantum Mpemba Effect from Exact Slow-Mode Selection in Constrained Rydberg Chains}

\author{Mingdi Xu}
\affiliation{School of Physics, Nankai University, Tianjin 300071, China}
\author{Kaixiang Lu}
\affiliation{School of Physics, Nankai University, Tianjin 300071, China}

\author{Zijun Wei}
\affiliation{School of Physics, Nankai University, Tianjin 300071, China}
\author{Xiang-Ping Jiang}
\affiliation{School of Physics, Hangzhou Normal University, Hangzhou, Zhejiang 311121, China}

\author{Haiping Hu}
\email{hhu@iphy.ac.cn}
\affiliation{Beijing National Laboratory for Condensed Matter Physics, Institute of Physics, Chinese Academy of Sciences, Beijing 100190, China}
\affiliation{School of Physical Sciences, University of Chinese Academy of Sciences, Beijing 100049, China}

\author{Lei Pan}%
\email{panlei@nankai.edu.cn}
\affiliation{School of Physics, Nankai University, Tianjin 300071, China}

\date{\today}

\begin{abstract}
Strong quantum Mpemba acceleration requires suppressing the slowest
visible Liouvillian relaxation channel, but a robust many-body mechanism
for enforcing such suppression remains challenging. We identify such a
mechanism in locally dephased constrained Rydberg chains through exact
slow-mode selection. For constrained single-spin-flip Hamiltonians, local
dephasing turns the Hamiltonian itself into an exact left Liouvillian
eigenmode, $\mathcal L^\dagger(H)=-\gamma H$. A finite-temperature
reference state generically overlaps with this $H$-like slow mode, whereas
translationally invariant states with $\mathrm{Tr}(H\rho_0)=0$ remove it and
are confined to the $Q=0$ operator sector. When the next visible $Q=0$ mode
decays faster, these selected states exhibit a strong quantum Mpemba effect.
We demonstrate this mechanism in the PXP chain for a zero-energy scar
eigenstate, the all-zero product state, and a translation-invariant $Z_2$
cat state, and show that it persists in the $(2,3)$ model and the
longer-range blockade family. Our results identify Liouvillian mode
visibility, rather than special scar wave functions, as the organizing
principle for anomalously fast relaxation in constrained open quantum
systems.
\end{abstract}

\maketitle

\textit{\color{blue} Introduction.---}
Nonequilibrium dynamics is a central theme in quantum many-body physics, driven by questions of thermalization, transport, relaxation, and information scrambling. Open quantum many-body systems form an especially rich class of nonequilibrium problems: coupling to an environment can destroy coherence, but it can also generate new dynamical structures. In recent years, dissipative many-body dynamics has attracted growing attention in engineered dissipation, dissipative quantum simulation, cold-atom experiments, and Rydberg-atom-array systems~\cite{Diehl2008,simulation,simulation1,simulation2,Exp1,Exp2,Exp3,Exp4,Exp5,Exp6,Exp7,Exp8,Exp9,Exp10,Exp_new1,Exp_new2,Exp_new3}. These studies have revealed anomalous relaxation phenomena beyond closed-system intuition~\cite{Pan2020,Ashida2017,Ashida2020,Bergholtz2021,Kawabata2023,Meden2023,Wang2024Absence,Gu2025Spontaneous,Lessa2025Strong,Sun2025Scheme,Mittal2026Fermion}.

The quantum Mpemba effect is a striking example of such anomalous relaxation: a state initially farther from stationarity can relax faster than a closer one. Its classical counterpart~\cite{ME,ME_classical1,ME_classical2,Inverse_ME1} has a long history and has been revisited in many nonequilibrium settings. Quantum versions have been explored in symmetry-restoration dynamics, integrable and localized systems, random circuits, and thermodynamic settings~\cite{QME1,QME101,QME2,QME3,QME4,QME_Exp1,QME_Exp2,QME18,QME20,QME21}. In parallel, open-system realizations have been studied in Markovian and non-Markovian relaxation, nonequilibrium reservoirs, dissipative many-body systems, localization and quasiperiodic platforms, bosonic and photonic setups, and trapped-ion experiments~\cite{Cha23,Str24,Cac24,Sha24,Lon24,Fur24,Lon24b,Lon24c,Don24,Wei25,Wes25,Ulv25,Liu25,Bag25,Cal25,Sal25,Li25,Das25,Lon25,Zho25,Son26,Wei26,Chi26,Lon26,Zha26,Bag26,Xia26b,Liu26b,Lon26b,Hong26Geometric,Chattopadhyay26Thermometry}. These developments already show that dissipative relaxation is not determined solely by the spectrum of decay rates: the initial state also controls which relaxation channels are accessed.

For Markovian dynamics, this statement is naturally formulated in terms of the Liouvillian superoperator, which replaces the Hamiltonian as the generator of time evolution~\cite{Lindblad1,Lindblad2,Breuer2002}. Its eigenvalues specify possible decay rates, but the observed relaxation also depends on mode visibility. A decay mode contributes only when the initial density matrix has nonzero overlap with the corresponding left Liouvillian eigenoperator~\cite{Prosen2008,Minganti2018,Talkington2024}. Thus the slowest Liouvillian mode need not control the relaxation; it can be completely invisible to a suitably selected initial state. In open quantum systems, the strong quantum Mpemba effect has a direct spectral interpretation: the farther state relaxes faster because it avoids the slowest decay channel visible to the reference state~\cite{OpenQME1,OpenQME2,OpenQME3,OpenQME4,OpenQME7,R1,R2}. The central many-body challenge is to make this invisibility robust: one needs
a physical principle that removes slow-mode visibility without fine tuning the
entire initial density matrix. 
Symmetry provides such a principle. In Lindblad dynamics, symmetries act directly in operator space, decompose the Liouvillian into independent sectors, constrain mode visibility, and may generate conserved quantities, multiple steady states, or long-lived coherent dynamics~\cite{BucaProsen2012,AlbertJiang2014,SanchezMunoz2019,Buca2019}. This suggests a strategy for realizing a robust strong quantum Mpemba effect: identify a slow left eigenmode visible to a natural reference state, and then find accessible initial states whose symmetry properties remove this mode while excluding competing slow channels in other sectors.

Here we show that locally dephased constrained Rydberg chains realize this strategy through an exact slow-mode selection rule. Rydberg blockade naturally generates constrained single-spin-flip dynamics, including the PXP model and longer-range blockade generalizations, and provides a programmable platform for constrained many-body dynamics and scars~\cite{Bernien2017,Turner2018,Browaeys2020,Serbyn2021,Lesanovsky2011,PanZhai2022,ChengLi2023}. Dephasing and dissipation are also natural ingredients in Rydberg platforms~\cite{LesanovskyGarrahan2013,Marcuzzi2014,Levi2016,Begoc2025}. We further test the mechanism in the $(2,3)$ model~\cite{Desaules2022Hypergrid}, which is a weakened-constraint graph distinct from the $R=2$ blockade model PPXPP. The central point is that the acceleration is not tied to the detailed wave function of a scar state, but follows from local dephasing and the single-spin-flip structure of the constrained Hamiltonian. We derive this rule in the PXP chain and then show that it persists across different constrained Hilbert-space graphs.
Recent protocol-based approaches, such as temporary reset operations, have also highlighted slow-mode visibility as a route to accelerated relaxation~\cite{Bao25Reset}. Our work identifies a complementary mechanism in which this visibility is removed by an exact symmetry-protected selection rule.

\textit{\color{blue} Exact selection and PXP evidence.---}
We first derive the selection rule and test it in the standard PXP chain. Under local dephasing, every coherence between two computational configurations decays at a rate proportional to their Hamming distance. Since the constrained Hamiltonians considered here contain only single-spin-flip matrix elements, all matrix elements of $H$ are single-flip coherences. As a result, the Hilbert-Schmidt adjoint Liouvillian satisfies the exact identity
\begin{equation}
	\label{eq:Hslow}
	\mathcal L^\dagger(H)=-\gamma H .
\end{equation}
The Hamiltonian itself is therefore an exact left Liouvillian slow mode. Its visibility is controlled by $\tr(H\rho_0)$: a finite-temperature state generically excites this channel, while an initial state with $\tr(H\rho_0)=0$ removes it.

This energy-selection condition must be supplemented by a symmetry filter. Slow modes may exist in operator-momentum sectors other than $Q=0$, and they would become visible to a generic non-translationally invariant density matrix. We therefore impose translation invariance, which confines the initial density matrix to the $Q=0$ operator sector. The resulting selection rule is
\begin{equation}
	\label{eq:selection-main}
	\tr(H\rho_0)=0,\qquad T\rho_0T^\dagger=\rho_0 .
\end{equation}
The first condition removes the exact $H$-like slow channel, while the second excludes all nonzero-$Q$ modes from the visible spectrum.

\begin{figure*}[t]
	\centering
	\includegraphics[width=0.98\textwidth]{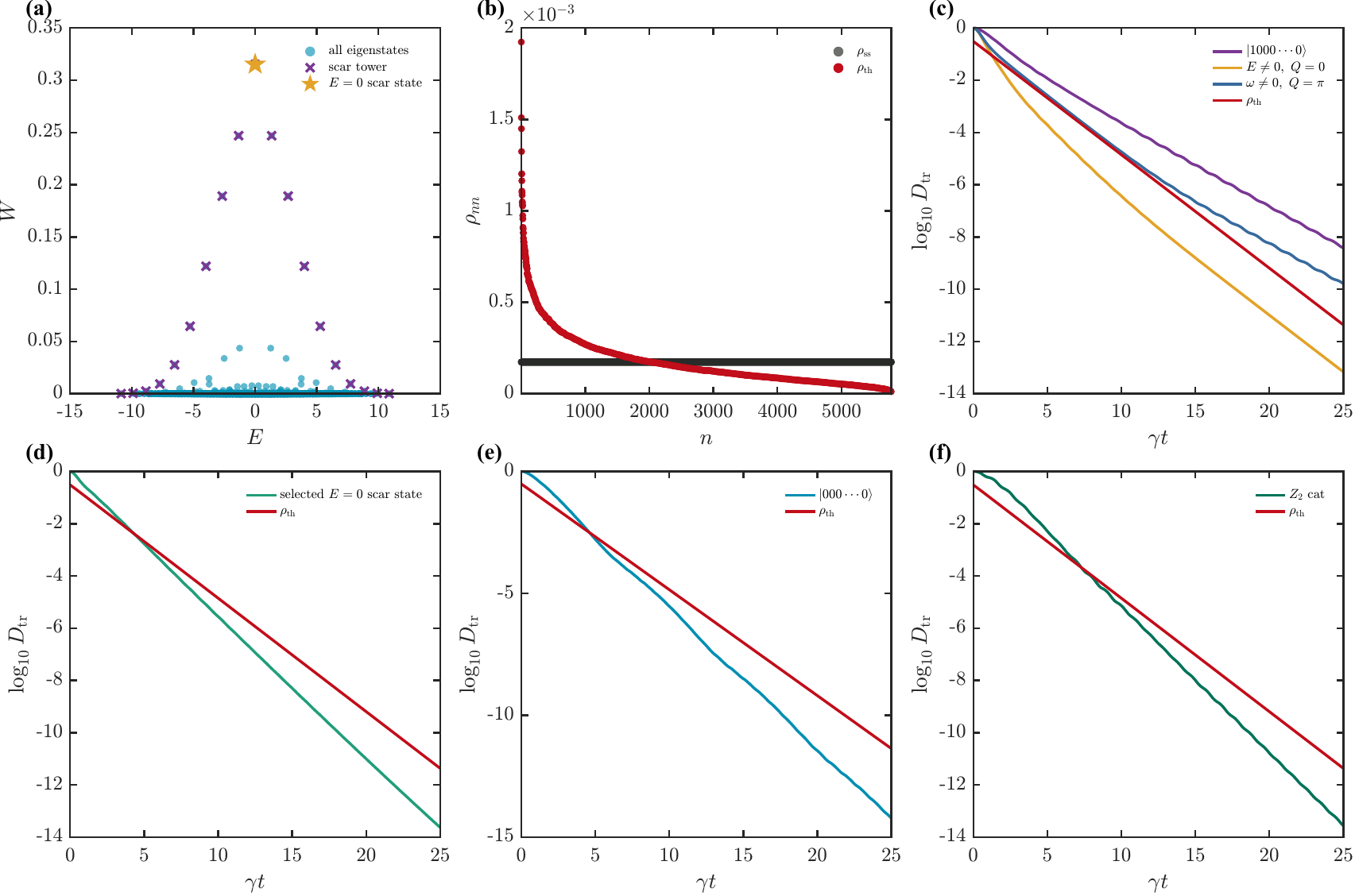}
	\caption{Strong quantum Mpemba effect in the locally dephased PXP chain. Periodic boundary conditions, $L=18$, $D=5778$, $\gamma=0.4$, and $\beta=0.25$ are used. (a) Spectral weight $W_n$ used to identify the representative zero-energy scar eigenstate. (b) Energy-basis diagonal weights of $\rhoSS$ and $\rhoTh$. (c) Control states violating one of the two selection conditions in Eq.~\eqref{eq:selection-main}: a non-translationally invariant product state can see nonzero-$Q$ modes, whereas a $Q=0$ state with nonzero energy retains the $H$-like channel. (d--f) Three selected states satisfying Eq.~\eqref{eq:selection-main} relax faster asymptotically than the same thermal reference.}
	\label{fig:fig1}
\end{figure*}

Figure~\ref{fig:fig1} demonstrates this mechanism in the locally dephased PXP chain. The steady state is the infinite-temperature state $\rhoSS=I_D/D$, and the reference state is the thermal ensemble $\rhoTh=e^{-\beta H}/Z$, which retains the $H$-like channel because $\tr(H\rhoTh)\neq0$. To construct one representative selected state, we identify a zero-energy scar eigenstate through its overlap with the two $Z_2$ product states,
$W_n=\max_{\pm}|\langle Z_2^{\pm}|E_n\rangle|^2$, where
$|Z_2^{\pm}\rangle=(|1010\cdots\rangle\pm|0101\cdots\rangle)/\sqrt{2}$.
This scar eigenstate is used as one physically meaningful example satisfying Eq.~\eqref{eq:selection-main}, not as the microscopic origin of the effect.
The control states in Fig.~\ref{fig:fig1}(c) illustrate the necessity of both filters. A non-translationally invariant product state can couple to nonzero-$Q$ slow modes, while a state with nonzero energy expectation fails to remove the $H$-like visible channel. Thus Eq.~\eqref{eq:selection-main} acts as a dynamical selection rule, rather than merely a convenient label for special initial states.

The main result is shown in Figs.~\ref{fig:fig1}(d)--\ref{fig:fig1}(f). Three distinct initial states satisfying Eq.~\eqref{eq:selection-main}---a selected zero-energy scar eigenstate, the $|0\cdots0\rangle$ product state, and the translation-invariant $Z_2$ cat state---all relax asymptotically faster than the same thermal reference in the trace distance
\begin{equation}
	\Dtr(\rho,\rho_{\rm ss})=\frac{1}{2}\|\rho-\rho_{\rm ss}\|_1,
	\label{eq:trace-distance}
\end{equation}
where $\|A\|_1=\tr\sqrt{A^\dagger A}$. Since the Liouvillian and the steady state are fixed throughout the comparison, the crossing is not caused by a trivial ordering of initial distances. Instead, the thermal state retains the exact $H$-like slow mode, whereas the selected states remove it and decay through a faster visible channel.

\textit{\color{blue} Constrained families.---}
We next show that the acceleration is not tied to the special scar structure of the standard PXP chain. The exact identity in Eq.~\eqref{eq:Hslow} follows only from the constrained single-spin-flip structure of the Hamiltonian, and should therefore survive changes of the constraint graph. We test this expectation in two types of constrained models: longer-range $R$-blockade models, where excitations are forbidden within distance $R$, and the $(2,3)$ model~\cite{Desaules2022Hypergrid}, where only three consecutive excitations are forbidden. The latter is a weakened-constraint model and should not be confused with the $R=2$ blockade model PPXPP. For these models, we use translation-symmetric density-wave cat states,

\begin{equation}
	\label{eq:zcat}
	\ket{Z_q^{\rm cat}}
	=
	\mathcal N_q
	\sum_{r=0}^{q-1}T^r
	\ket{
		\left(\underbrace{10\cdots0}_{q}\right)
		\left(\underbrace{10\cdots0}_{q}\right)
		\cdots
	} .
\end{equation}
Each block $\underbrace{10\cdots0}_{q}$ contains one excitation followed by $q-1$ zeros; for example, $q=2,3,4,5$ correspond to the density-wave patterns $1010\cdots$, $100100\cdots$, $10001000\cdots$, and $1000010000\cdots$. These cat-state density matrices are translationally invariant and hence lie in the $Q=0$ operator sector. Since the Hamiltonians contain only off-diagonal single-spin-flip matrix elements, they also satisfy $\tr(H\rho_0)=0$. They therefore obey the same selection rule as Eq.~\eqref{eq:selection-main}.

\begin{figure*}[t]
	\centering
	\includegraphics[width=0.96\textwidth]{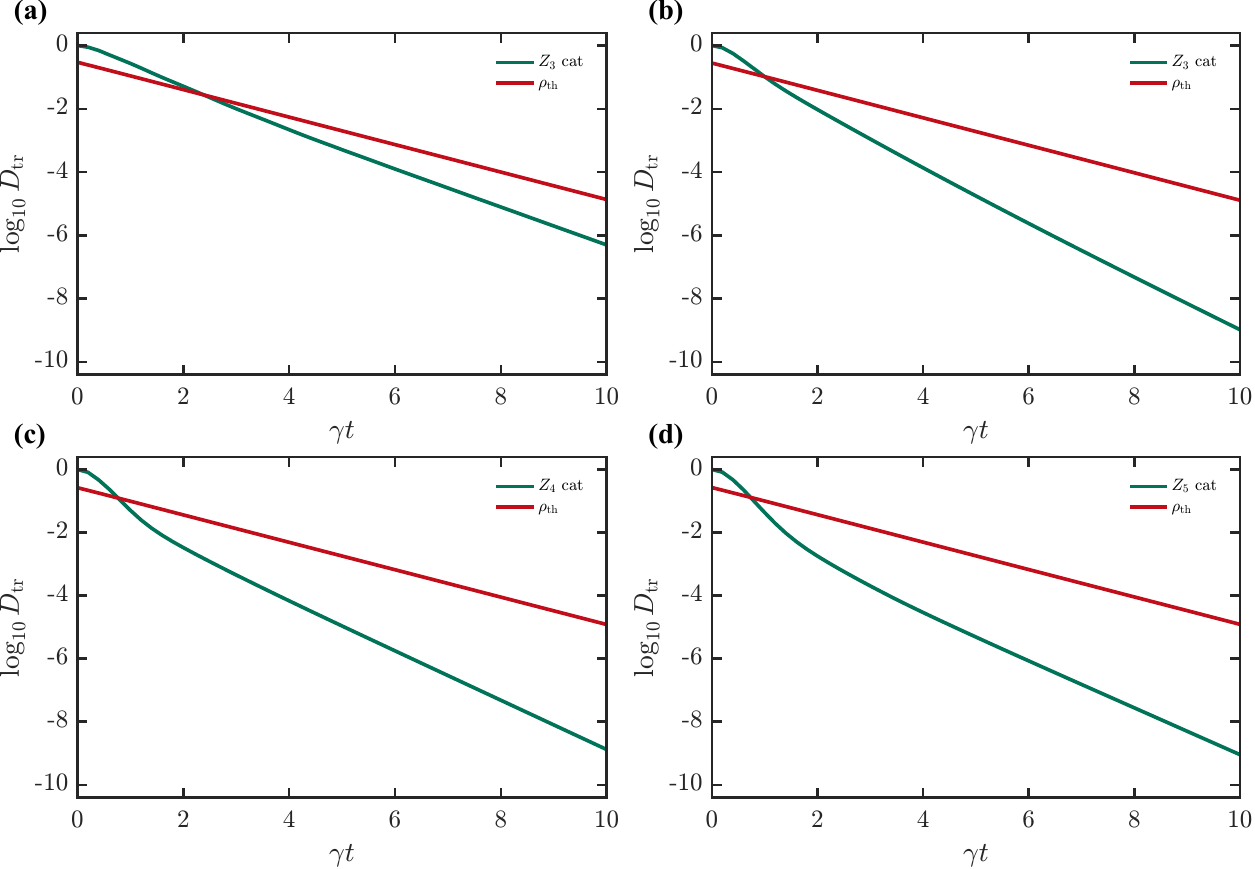}
	\caption{Cross-model SQME from the same selection rule. Each panel compares a translation-invariant cat state with $\tr(H\rho_0)=0$ to the thermal reference $\rhoTh$. Periodic boundary conditions, $\gamma=0.1$, $\beta=0.25$, and $\gamma t\le10$ are used. The models are (a) the $(2,3)$ model, $L=12,D=1499$; (b) PPXPP, $L=21,D=3063$; (c) PPPXPPP, $L=24,D=2287$; and (d) PPPPXPPPP, $L=30,D=4612$.}
	\label{fig:fig2}
\end{figure*}

Figure~\ref{fig:fig2} confirms this prediction across several constrained models. In the $(2,3)$ model and in longer-range blockade variants, the selected cat states relax asymptotically faster than the same thermal reference. Changing the constraint modifies the Hilbert-space graph and the numerical decay rates, but it does not change the visibility classification by $\tr(H\rho_0)$ and operator momentum. This cross-model agreement shows that the strong quantum Mpemba effect is controlled by exact slow-mode selection, rather than by the detailed wave function of a particular PXP scar state.

\textit{\color{blue} Slow-mode selection.---}
We now give a mode-resolved interpretation of the acceleration. Let
$\Liouv(R_\alpha)=\lambda_\alpha R_\alpha$,
$\Liouv^\dagger(L_\alpha)=\lambda_\alpha^*L_\alpha$, and
$\tr(L_\alpha^\dagger R_\beta)=\delta_{\alpha\beta}$. The relaxation from an initial state can be expanded as
\begin{equation}
	\rho(t)-\rhoSS
	=
	\sum_\alpha c_\alpha e^{\lambda_\alpha t}R_\alpha,
	\qquad
	c_\alpha=
	\tr[L_\alpha^\dagger(\rho_0-\rhoSS)] .
	\label{eq:mode-expansion}
\end{equation}
Thus the observed long-time decay is controlled not by the slowest Liouvillian eigenvalue alone, but by the slowest mode with nonzero visibility coefficient $c_\alpha$.
The exact identity in Eq.~\eqref{eq:Hslow} identifies one such visible channel. Since $H$ is a left eigenoperator with decay rate $\gamma$, its coefficient is proportional to
\begin{equation}
	c_H(\rho_0)
	\propto
	\tr[H(\rho_0-\rhoSS)]
	=
	\tr(H\rho_0),
	\label{eq:cH}
\end{equation}
where $\rhoSS=I_D/D$ and $\tr H=0$. A finite-temperature state generically has $c_H\neq0$ and therefore excites the $H$-like slow mode. By contrast, states satisfying $\tr(H\rho_0)=0$ remove this exact channel. This is the first part of the selection rule.

The second part is imposed by translation symmetry. The operator-space translation superoperator $\Tcal(O)=TOT^\dagger$ decomposes the Liouvillian into momentum sectors. A translationally invariant density matrix lies entirely in the $Q=0$ sector and is orthogonal to all $Q\neq0$ left modes. Since $[H,T]=0$, both the thermal reference and the selected states are restricted to $Q=0$, but only the selected states remove the $H$-like mode. Their asymptotic decay must therefore proceed through the next visible non-$H$ mode in the $Q=0$ sector, with decay rate denoted by $r_{\rm next}^{Q=0}$.
This gives a simple spectral criterion for the strong quantum Mpemba effect. We define
\begin{equation}
	\eta=\frac{r_{\rm fast}}{r_{\rm th}},
	\label{eq:eta}
\end{equation}
where $r_{\rm th}$ is the visible decay rate of the thermal reference and $r_{\rm fast}$ is that of the selected state. In the cases studied here, the thermal tail is governed by the exact $H$-like mode, so $r_{\rm th}=\gamma$, while the selected state decays with $r_{\rm fast}=r_{\rm next}^{Q=0}$. Hence $\eta>1$ directly signals that removing the exact slow channel produces faster asymptotic relaxation. 

\begin{figure}[t]
	\centering
	\includegraphics[width=0.94\columnwidth]{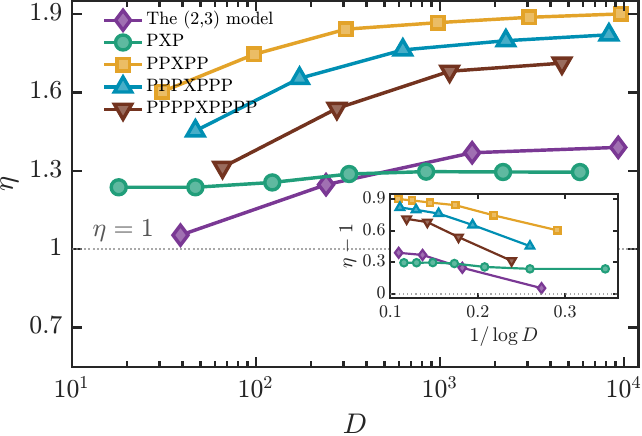}
	\caption{Visible-rate ratio $\eta=r_{\rm fast}/r_{\rm th}$ versus constrained Hilbert-space dimension $D$. The $(2,3)$ model is not an $R$-blockade member. All data use periodic boundary conditions, local dephasing, $\gamma=0.1$, and $\beta=0.25$. Values above unity indicate
	asymptotically faster relaxation than the thermal reference. The inset
	plots $\eta-1$ against $1/\log D$ and shows an apparent saturation trend,
	supporting the persistence of a finite selected-state advantage over the
	accessible Hilbert-space dimensions.}
	\label{fig:fig3}
\end{figure}

Figure~\ref{fig:fig3} shows that $\eta$ remains above unity over the accessible size range for the standard PXP chain, the $(2,3)$ model, and the longer-range blockade family. While these data do not by themselves constitute a rigorous thermodynamic-limit proof, the finite-size trend and the inset of Fig.~\ref{fig:fig3} show an apparent saturation of $\eta-1$, supporting the persistence of a finite selected-state advantage over the accessible Hilbert-space dimensions. Interestingly, recent work has shown that generic initial states in certain open quantum systems can exhibit a typical strong quantum Mpemba effect in the thermodynamic limit, where the overlap with the slowest relaxation mode becomes vanishingly small for most initial states~\cite{Bao26Typicality}. Although the mechanism is different from the exact symmetry-protected selection rule studied here, both results highlight the importance of initial-state visibility of slow Liouvillian modes for relaxation dynamics.

\begin{figure}[t]
	\centering
	\includegraphics[width=0.95\columnwidth]{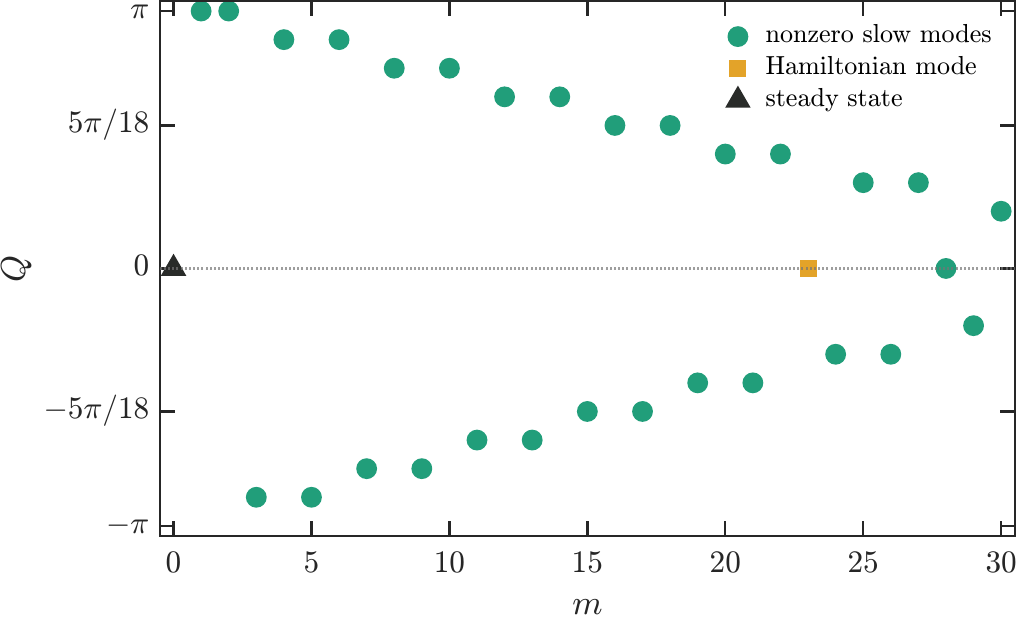}
	\caption{Operator momenta of slow Liouvillian modes in the locally dephased PXP chain with $L=18$ and $\gamma=0.4$. Translationally invariant density matrices couple only to $Q=0$ modes. The $H$-like mode has rate $r_H=\gamma$, while the next visible non-$H$ mode in the $Q=0$ sector has a larger decay rate. Imposing $\tr(H\rho_0)=0$ removes the $H$-like mode and leaves the faster visible channel.}
	\label{fig:fig4}
\end{figure}

Figure~\ref{fig:fig4} provides a direct mode-resolved check in the standard PXP chain. Some of the globally slow nonsteady modes lie in nonzero-$Q$ sectors and are therefore invisible to translationally invariant density matrices. Within the visible $Q=0$ sector, the exact $H$-like mode is the slowest thermal channel, while the next non-$H$ mode decays faster. This explains why the thermal state has a slower tail, whereas states satisfying Eq.~\eqref{eq:selection-main} bypass this tail and relax through the faster $Q=0$ channel. The spectral structure in Fig.~\ref{fig:fig4} is therefore the mode-space origin of the asymptotically faster relaxation shown in Figs.~\ref{fig:fig1} and~\ref{fig:fig2}.

Combining these observations yields a sufficient long-time condition for the strong quantum Mpemba effect. Suppose that the thermal reference is
translationally invariant and has a nonzero $H$-mode visibility, $c_H(\rho_{\rm th})\neq0$, so that its long-time tail is governed by the $H$-like mode with rate $\gamma$. If a translationally invariant state $\rho_A$ satisfies $\mathrm{Tr}(H\rho_A)=0$ and the next visible non-$H$ mode in the $Q=0$ sector obeys $r_{\rm next}^{Q=0}>\gamma$, then
\begin{equation}
	\frac{\Dtr(\rho_A(t),\rhoSS)}
	{\Dtr(\rhoTh(t),\rhoSS)}
	\sim
	C\exp[-(r_{\rm next}^{Q=0}-\gamma)t]
	\rightarrow 0,
	\qquad t\rightarrow\infty ,
	\label{eq:sqme-condition}
\end{equation}
Here $C$ is a nonuniversal prefactor determined by visible mode amplitudes and right-eigenoperator norms. If degeneracies occur near $\lambda=-\gamma$, the condition should be understood as the vanishing of the projection onto the corresponding left eigenspace. Details on thermal overlaps, residual checks, fitting procedures, data tables, and $Z_q$ cat-state preparation are given in the Supplemental Material.

\textit{\color{blue} Conclusion and Outlook.---}
We have shown that locally dephased constrained Rydberg chains exhibit a strong quantum Mpemba effect through an exact Liouvillian mode-selection mechanism. The central observation is that the constrained single-spin-flip structure makes the Hamiltonian itself an exact left slow mode of the dephased Liouvillian, $\Liouv^\dagger(H)=-\gamma H$. A thermal reference state generically retains this channel, whereas translationally invariant states with vanishing energy expectation remove it and are confined to the $Q=0$ operator sector. When the next visible $Q=0$ mode decays faster than this exact $H$-like channel, the selected states relax asymptotically faster even though they start farther from the steady state.

This mechanism is directly relevant to programmable Rydberg platforms. PXP and longer-range blockade Hamiltonians can be realized in Rydberg atom arrays~\cite{Lesanovsky2011,Bernien2017,Browaeys2020,PanZhai2022,ChengLi2023}, while dephasing can be introduced through laser phase noise, measurement-induced channels, or engineered dissipative processes~\cite{LesanovskyGarrahan2013,Marcuzzi2014,Levi2016,Begoc2025}. Importantly, the effect does not require preparing an exact scar eigenstate or reconstructing the full density matrix. Simple product states, translation-symmetrized density-wave states, or shot-averaged translated preparations already provide experimentally accessible ways to implement the selection rule. Local densities, imbalance-type observables, structure factors, or shadow-estimated distances can then be used to detect the change of long-time decay rate.

Our results recast the strong quantum Mpemba effect as a problem of Liouvillian mode accessibility. Quantum many-body scars provide a useful preparation route, but they are not the origin of the acceleration: the mechanism is fixed by exact slow-mode selection and survives beyond the scarred PXP setting. The acceleration does not rely on fine tuning or on the detailed wave function of a particular scar state; it follows from an exact left mode together with a symmetry-imposed visibility filter. This viewpoint suggests a broader route for engineering anomalously fast relaxation in constrained, symmetric, and experimentally controllable open quantum many-body systems.\\

\textit{\color{blue} Acknowledgements.---}
The work is supported by the National Key Research and Development Program of China (Grants No. 2022YFA1405800 and No. 2023YFA1406704) and National Natural Science Foundation of China (Grant
No. 12304290 and No. 12474496). LP also acknowledges support from the Fundamental Research Funds for the Central Universities. \\

\bibliography{Reference}

\clearpage
\onecolumngrid
\section*{Supplemental Material}

\setcounter{equation}{0}
\renewcommand{\theequation}{S\arabic{equation}}
\renewcommand{\theHequation}{S\arabic{equation}}

\setcounter{table}{0}
\renewcommand{\thetable}{S\arabic{table}}
\renewcommand{\theHtable}{S\arabic{table}}

\setcounter{figure}{0}
\renewcommand{\thefigure}{S\arabic{figure}}
\renewcommand{\theHfigure}{S\arabic{figure}}

\section{From the microscopic Rydberg model to constrained single-flip Hamiltonians}

We first recall how the constrained single-flip Hamiltonians used in the main text arise from a microscopic Rydberg atom chain. In the rotating frame, the Hamiltonian of a one-dimensional Rydberg array is
\begin{equation}
	H_{\rm Ry}
	=
	\sum_{j=1}^{L}
	\left(
	\frac{\Omega}{2}X_j-\Delta n_j
	\right)
	+
	\sum_{i<j}V_{ij}n_i n_j ,
	\qquad
	V_{ij}=\frac{C_6}{|r_i-r_j|^6}.
	\label{eq:HRy_supp}
\end{equation}
Here $X_j$ flips the atom between the ground and Rydberg states, $n_j=|1\rangle_j\langle 1|$ is the Rydberg occupation, and $P_j=1-n_j$ projects onto the ground state. We consider the resonant case $\Delta=0$ and focus on the hard-blockade regime, where configurations with two Rydberg excitations separated by $d=1,\ldots,R$ lattice spacings are projected out. The corresponding constrained subspace is defined by
\begin{equation}
	n_j n_{j+d}=0,
	\qquad
	d=1,\ldots,R ,
	\label{eq:blockade_constraint_supp}
\end{equation}
with periodic boundary conditions understood unless otherwise stated.

Following the standard degenerate-perturbation or Schrieffer-Wolff projection argument for the PXP model, we introduce the projector onto the $R$-blockade subspace,
\begin{equation}
	\mathcal P_R
	=
	\prod_j
	\prod_{d=1}^{R}
	\left(1-n_j n_{j+d}\right).
	\label{eq:PR_projector_supp}
\end{equation}
In the strong-blockade limit, the interaction term in Eq.~\eqref{eq:HRy_supp} only enforces the constraint. The leading nontrivial dynamics within the constrained manifold is therefore obtained by projecting the Rabi drive,
\begin{equation}
	H_{\rm eff}^{(1)}
	=
	\mathcal P_R
	\left(
	\frac{\Omega}{2}\sum_j X_j
	\right)
	\mathcal P_R .
	\label{eq:Heff_projected_supp}
\end{equation}
Within the constrained Hilbert space, the projected spin flip at site $j$ is allowed only when all sites within blockade distance $R$ on both sides are unexcited. Thus,
\begin{equation}
	\mathcal P_R X_j \mathcal P_R
	=
	\left(
	\prod_{a=1}^{R}P_{j-a}
	\right)
	X_j
	\left(
	\prod_{a=1}^{R}P_{j+a}
	\right).
	\label{eq:projected_flip_supp}
\end{equation}
After setting the overall energy scale $\Omega/2$ to unity, the arbitrary-radius blockade Hamiltonian is
\begin{equation}
	H_R
	=
	\sum_{j=1}^{L}
	\left(
	\prod_{a=1}^{R}P_{j-a}
	\right)
	X_j
	\left(
	\prod_{a=1}^{R}P_{j+a}
	\right).
	\label{eq:HR_blockade_supp}
\end{equation}
For $R=1$, Eq.~\eqref{eq:HR_blockade_supp} reduces to the standard PXP Hamiltonian,
\begin{equation}
	H_{R=1}
	=
	\sum_j P_{j-1}X_jP_{j+1}.
\end{equation}
For $R=2,3,4$, it gives the PPXPP, PPPXPPP, and PPPPXPPPP models, respectively.

The resonant projected Hamiltonian in Eq.~\eqref{eq:HR_blockade_supp} is the minimal $R$-blockade model studied in the main text. Detuning terms and residual long-range diagonal interaction tails are not included in this minimal Hamiltonian. They should be viewed as experimental perturbations to the ideal constrained single-flip model. This distinction is important because the exact identity $\mathcal L^\dagger(H_R)=-\gamma H_R$ relies on the fact that $H_R$ contains only off-diagonal single-spin-flip matrix elements in the constrained computational basis.

We also study the $(2,3)$ model, which is conceptually distinct from the $R$-blockade sequence. Following Ref.~\cite{Desaules2022Hypergrid}, it belongs to a weakened-constraint family usually denoted as the $(k,k+1)$ models. In this family, the constraint is that every block of $k+1$ consecutive sites contains at most $k$ excitations. Equivalently, one introduces a projector $\mathcal P_{(k,k+1)}$ that removes all configurations containing $k+1$ consecutive excitations, and defines
\begin{equation}
	H_{(k,k+1)}
	=
	\mathcal P_{(k,k+1)}
	\left(
	\sum_j X_j
	\right)
	\mathcal P_{(k,k+1)} .
	\label{eq:Hkk1_supp}
\end{equation}
The case $k=1$ gives the PXP model, because it forbids the pattern $11$. Increasing $k$ weakens the constraint and interpolates toward the free spin-$1/2$ model.
The $(2,3)$ model is the $k=2$ member of this weakened-constraint family. It forbids only three consecutive excitations,
\begin{equation}
	n_j n_{j+1} n_{j+2}=0 ,
	\label{eq:23_constraint_supp}
\end{equation}
but allows adjacent pairs of excitations. Its Hamiltonian is
\begin{equation}
	H_{(2,3)}
	=
	\mathcal P_{(2,3)}
	\left(
	\sum_j X_j
	\right)
	\mathcal P_{(2,3)} .
	\label{eq:H23_supp}
\end{equation}
Thus the $(2,3)$ model should not be identified with the $R=2$ blockade model PPXPP. The latter forbids any two excitations within distance two, whereas the $(2,3)$ model only forbids the local pattern $111$.

\subsection{Hilbert-space dimensions under periodic boundary conditions}

For periodic boundary conditions, the constrained Hilbert-space dimension is the number of binary strings of length $L$ satisfying the corresponding blockade rule on a ring. For the $R$-blockade family, this means that two excitations cannot be separated by $d=1,\ldots,R$ lattice spacings. Equivalently,
\begin{equation}
	n_j n_{j+d}=0,
	\qquad
	d=1,\ldots,R ,
\end{equation}
with all site indices understood modulo $L$.

A convenient way to compute the periodic-chain dimension is through a transfer matrix. Let $\boldsymbol\sigma=(\sigma_1,\ldots,\sigma_R)$ and $\boldsymbol\tau=(\tau_1,\ldots,\tau_R)$ be binary words of length $R$. The transfer matrix for the $R$-blockade constraint is
\begin{equation}
	\left(\mathbb A_R\right)_{\boldsymbol\sigma,\boldsymbol\tau}=\left(\prod_{a=1}^{R-1}\delta_{\tau_a,\sigma_{a+1}}\right)
	\prod_{a=1}^{R}\left(1-\sigma_a\tau_R\right).
	\label{eq:transfer_R_supp}
\end{equation}
The first factor shifts the memory window by one site, while the second factor forbids adding a new excitation within distance $R$ of any previous excitation. The Hilbert-space dimension for a periodic chain is then
\begin{equation}
	D_R^{\rm PBC}(L)=\mathrm{Tr}\,\mathbb A_R^L .
	\label{eq:dim_R_pbc_supp}
\end{equation}
This trace automatically enforces the constraint across the boundary. For $R=1$, Eq.~\eqref{eq:dim_R_pbc_supp} reduces to the usual PXP result
\begin{equation}
	D_{R=1}^{\rm PBC}(L)=F_{L-1}+F_{L+1},
\end{equation}
where $F_L$ is the Fibonacci number.

The $(2,3)$ model used in the main text is different from the $R$-blockade sequence. It forbids three consecutive excitations but allows adjacent pairs. Its periodic-chain dimension can be computed from a two-bit transfer matrix,
\begin{equation}
	\left(\mathbb B\right)_{\boldsymbol\sigma,\boldsymbol\tau}=	\delta_{\tau_1,\sigma_2}
	\left(1-\sigma_1\sigma_2\tau_2\right),
	\qquad
	\boldsymbol\sigma=(\sigma_1,\sigma_2),\quad
	\boldsymbol\tau=(\tau_1,\tau_2),
	\label{eq:transfer_23_supp}
\end{equation}
so that
\begin{equation}
	D_{(2,3)}^{\rm PBC}(L)=\mathrm{Tr}\,\mathbb B^L .
	\label{eq:dim_23_pbc_supp}
\end{equation}

For reference, the periodic-chain Hilbert-space dimensions used in the main-text figures are listed in Table~\ref{tab:hilbert_dimensions_supp}.

\begin{table}[t]
	\caption{Largest system sizes and corresponding constrained Hilbert-space dimensions reached in our numerical calculations under periodic boundary conditions.}
	\label{tab:hilbert_dimensions_supp}
	\begin{ruledtabular}
		\begin{tabular}{l c c}
			model & $L_{\max}$ & $D^{\rm PBC}$ \tabularnewline
			PXP ($R=1$ blockade) & $18$ & $5778$ \tabularnewline
			PPXPP ($R=2$ blockade) & $24$ & $9642$ \tabularnewline
			PPPXPPP ($R=3$ blockade) & $28$ & $8300$ \tabularnewline
			PPPPXPPPP ($R=4$ blockade) & $30$ & $4612$ \tabularnewline
			$(2,3)$ model, no $111$ strings & $15$ & $9327$ \tabularnewline
		\end{tabular}
	\end{ruledtabular}
\end{table}

\section{Numerical convention and definition of the rate ratio}

All time-evolution and Liouvillian-spectrum calculations use the same local-dephasing convention,
\begin{equation}
	\Liouv(\rho)=-i[H,\rho]
	+\gamma\sum_j\left(2n_j\rho n_j-\{n_j,\rho\}\right).
\end{equation}
In the computational basis,
\begin{equation}
	\Diss(\ket{\mu}\bra{\nu})
	=-\gamma d(\mu,\nu)\ket{\mu}\bra{\nu},
\end{equation}
where \(d(\mu,\nu)\) is the Hamming distance between the two binary configurations. With this convention, the exact \(H\)-like left slow mode of a single-flip Hamiltonian has decay rate \(r_H=\gamma\).

The ratio reported in Fig.~3 of the main text is
\begin{equation}
	\eta=\frac{r_{\rm fast}}{r_{\rm th}},
\end{equation}
where the two rates are defined by the slowest visible Liouvillian modes for the candidate fast state and the thermal state. In time-evolution plots, if the horizontal axis is \(\gamma t\), the long-time straight-line slope of \(\log_{10}\Dtr\) obeys
\begin{equation}
	s=-\frac{r}{\gamma\ln 10}.
\end{equation}
Therefore the rate ratio inferred from the long-time tail agrees with the ratio inferred from the visible Liouvillian spectrum. In practice we fit only before the numerical plateau and use spectral ordering to verify that the fitted tail corresponds to a single visible slow mode.

\section{Exact \texorpdfstring{$H$-like}{H-like} left slow mode}

We now show explicitly why the Hamiltonian is an exact left Liouvillian eigenoperator for the minimal constrained single-flip models. Let $E_{\mu\nu}=\ket{\mu}\bra{\nu}$ denote a computational-basis operator. A constrained single-spin-flip Hamiltonian has nonzero matrix elements only between configurations that differ by one spin flip,
\begin{equation}
	H=\sum_{\langle\mu,\nu\rangle}H_{\mu\nu}\ket{\mu}\bra{\nu},
	\qquad d(\mu,\nu)=1 .
\end{equation}
For the local-dephasing convention used throughout the work,
\begin{equation}
	\Liouv^\dagger(O)=i[H,O]
	+\gamma\sum_j\left(2n_j O n_j-\{n_j,O\}\right).
\end{equation}
The coherent contribution vanishes for $O=H$, since $i[H,H]=0$. The dissipative contribution acts diagonally in the computational operator basis: each single-flip coherence has Hamming distance one and therefore decays at rate $\gamma$. Hence
\begin{equation}
	\Diss^\dagger(H)=-\gamma H,
	\qquad
	\Liouv^\dagger(H)=-\gamma H .
\end{equation}
This identity is exact for the projected single-flip Hamiltonian and local dephasing. Diagonal detuning terms, residual interaction tails, or off-graph couplings would be perturbations away from this minimal fixed point and are therefore not included in the analytic proof.

The visibility of this left mode in an initial deviation $\Delta\rho_0=\rho_0-\rhoSS$ is
\begin{equation}
	c_H(\rho_0)\propto
	\tr(H\Delta\rho_0)
	=\tr(H\rho_0),
\end{equation}
where $\rhoSS=I_D/D$ and $\tr H=0$. Thus every initial density matrix with $\tr(H\rho_0)=0$ strictly avoids the exact $H$-like slow channel.

A finite-temperature reference state,
\begin{equation}
	\rho_\beta=\frac{e^{-\beta H}}{\tr e^{-\beta H}},
\end{equation}
generically has a nonzero overlap with this channel,
\begin{equation}
	\tr(H\rho_\beta)=-\partial_\beta\log \tr e^{-\beta H}.
\end{equation}
The spectral symmetry used in this expression follows from the excitation-parity operator
\begin{equation}
	\Pi=(-1)^{N_{\rm exc}},\qquad N_{\rm exc}=\sum_j n_j .
\end{equation}
Since the minimal constrained Hamiltonians contain only single-spin-flip matrix elements, they change $N_{\rm exc}$ by one and hence anticommute with parity,
\begin{equation}
	\{\Pi,H\}=0 .
\end{equation}
Therefore, if $\ket{E}$ has energy $E\neq0$, then $\Pi\ket{E}$ has energy $-E$. The nonzero spectrum is thus symmetric under $E\leftrightarrow -E$. If $N_0$ denotes the number of zero modes, then
\begin{equation}
	\tr(H\rho_\beta)
	=-\frac{2\sum_{E>0}E\sinh(\beta E)}
	{N_0+2\sum_{E>0}\cosh(\beta E)}.
\end{equation}
Therefore, for any nontrivial constrained graph and fixed $\beta>0$, the thermal reference excites the exact $H$-like slow channel. In the high-temperature limit,
\begin{equation}
	\tr(H\rho_\beta)=-\beta\frac{\tr H^2}{D}+O(\beta^3),
\end{equation}
so the amplitude vanishes linearly as $\beta\to0$, while the associated
asymptotic rate remains $r_H=\gamma$ whenever this is the slowest visible
thermal mode.

\section{Density-matrix translation invariance and operator-momentum selection}

The symmetry condition used in the main text is formulated directly for the density matrix. With periodic boundary conditions, define
\begin{equation}
	\Tcal(O)=T O T^\dagger .
\end{equation}
Operator-space sectors are labeled by \(Q\) through
\begin{equation}
	\Tcal(O_Q)=e^{iQ}O_Q .
\end{equation}
A translationally invariant density matrix satisfies
\begin{equation}
	T\rho T^\dagger=\rho ,
\end{equation}
which is equivalent to saying that \(\rho\) belongs entirely to the \(Q=0\) operator sector. Equivalently, \(\rho\) is Hilbert-Schmidt orthogonal to every operator that has no \(Q=0\) component.

Because \([\Liouv,\Tcal]=0\) under periodic boundary conditions, Liouvillian eigenmodes can be block-diagonalized by \(Q\). Let \(L_Q\) be a left mode with operator momentum \(Q\), and let \(\Delta\rho=\rho_0-\rhoSS\) be translationally invariant. Then
\begin{align}
	C_Q&=\tr(L_Q^\dagger\Delta\rho)\\
	&=\tr[\Tcal(L_Q^\dagger\Delta\rho)]
	=e^{-iQ}C_Q .
\end{align}
For \(Q\neq0\), this implies \(C_Q=0\). Therefore translationally invariant initial density matrices cannot see nonzero-operator-momentum slow modes. This is the sense in which the main text classifies initial conditions by translation invariance, while using \(Q\) to label operator-space sectors and Liouvillian modes.

If the \(H\)-like mode is the slowest nonsteady mode in the visible \(Q=0\) sector and the next non-\(H\) mode satisfies
\begin{equation}
	r_{\rm next}^{Q=0}>\gamma,
\end{equation}
then any initial state obeying \(\Tcal(\rho_A)=\rho_A\) and \(\tr(H\rho_A)=0\) has
\begin{equation}
	\Dtr(\rho_A(t),\rhoSS)
	=O(e^{-r_{\rm next}^{Q=0}t}),
\end{equation}
whereas a finite-temperature state has
\begin{equation}
	\Dtr(\rho_\beta(t),\rhoSS)=\Theta(e^{-\gamma t}).
\end{equation}
Consequently,
\begin{equation}
	\frac{\Dtr(\rho_A(t),\rhoSS)}
	{\Dtr(\rho_\beta(t),\rhoSS)}
	\le C e^{-(r_{\rm next}^{Q=0}-\gamma)t}
	\rightarrow0 .
\end{equation}
This provides a sufficient long-time condition for the strong quantum Mpemba effect.

\section{Numerical implementation and stability criteria}

\subsection*{Projection to \(Q=0\)}

The rate ratios in Fig.~3 of the main text come from translationally invariant initial density matrices, so we first project the full operator space into the \(Q=0\) sector. Let \(E_{ab}=\ket{a}\bra{b}\) and \(\Tcal(E_{ab})=T E_{ab}T^\dagger\). If
\(\mathcal O=\{E_{ab},\Tcal(E_{ab}),\ldots,\Tcal^{\ell-1}(E_{ab})\}\)
is an orbit of length \(\ell\), the normalized \(Q=0\) basis vector is
\begin{equation}
	\ket{\mathcal O;Q=0}
	=\frac{1}{\sqrt{\ell}}\sum_{r=0}^{\ell-1}\Tcal^r(E_{ab}).
\end{equation}
Putting these orbit basis vectors into the columns of \(U_{Q=0}\) gives the projected Liouvillian
\begin{equation}
	\Liouv_{Q=0}=U_{Q=0}^\dagger \Liouv U_{Q=0}.
\end{equation}
For the rate-ratio scans in Fig.~3, we further project to the
\(P_{CT}=+1\), \(R=+1\) block,
\begin{equation}
	\Liouv_{++}=W_R^\dagger W_{P_{CT}}^\dagger
	\Liouv_{Q=0}
	W_{P_{CT}}W_R ,
\end{equation}
because both the thermal state and the candidate fast-state density matrices compared in the main text are translationally invariant and therefore lie in this visible subspace. In Fig.~4 of the main text, all \(Q\) sectors are retained, and the \(Q=0,\pi\) sectors are further decomposed by \(P_{CT}\) and \(R\).

\subsection*{Eigensolver parameters and convergence}

The \(Q=0\) rate-ratio calculations for Fig.~3 of the main text use sparse partial diagonalization and retain the slowest 16 modes,
\begin{equation}
	\texttt{eigsTarget}=\texttt{largestreal},\quad
	\texttt{tol}=10^{-10},\quad
	\texttt{maxit}=5000,
\end{equation}
with a Krylov subspace dimension \(p\ge 48\). If \texttt{largestreal} fails, the calculation falls back to \texttt{lr}. The blockwise \(L=18\) scan in Fig.~4 uses the stricter parameters
\begin{equation}
	\texttt{tol}=10^{-11},\qquad
	\texttt{maxit}=8000,
	\qquad
	p\ge 64 ,
\end{equation}
and retains the 8 slowest modes, ordered by \(\mathrm{Re}\,\lambda\), in each reduced symmetry block. In the latest \(L=18\) data set, the largest reduced block dimension is \(927870\).

\subsection*{Residuals, biorthogonal normalization, and visibility}

For a right eigenmode $R_\alpha$, numerical stability is monitored by the relative residual
\begin{equation}
	\varepsilon_\alpha=
	\frac{\|\Liouv R_\alpha-\lambda_\alpha R_\alpha\|_2}
	{\|R_\alpha\|_2}.
\end{equation}
In the largest matrix-free $Q=0$ calculations used in Fig.~\ref{fig:fig3}, representative residuals are of order $10^{-13}$. For example, the largest PXP point $(L,D,N_{Q=0})=(18,5778,1855100)$ gives an $H$-mode residual of $9.29\times10^{-14}$. The largest point of the $(2,3)$ model, $(15,9327,5799601)$, gives $\varepsilon_H=1.11\times10^{-13}$ and $\varepsilon_2=1.15\times10^{-13}$.

For left and right Liouvillian eigenmodes $\{L_\alpha,R_\alpha\}$, we use the biorthogonal convention
\begin{equation}
	\Liouv R_\alpha=\lambda_\alpha R_\alpha,
	\qquad
	\Liouv^\dagger L_\alpha=\lambda_\alpha^* L_\alpha,
	\qquad
	\tr(L_\alpha^\dagger R_\beta)=\delta_{\alpha\beta}.
\end{equation}
The visibility of mode $\alpha$ for an initial deviation $\Delta\rho_0=\rho_0-\rhoSS$ is
\begin{equation}
	c_\alpha=\tr(L_\alpha^\dagger\Delta\rho_0).
\end{equation}
Only modes with $c_\alpha\neq0$ enter the long-time tail. The two central invisibility statements in this work are analytic: translationally invariant density matrices are orthogonal to all $Q\neq0$ left modes, and states with $\tr(H\rho_0)=0$ are orthogonal to the exact $H$-like left mode. Numerical overlap thresholds are used only as diagnostics.

For diagnostic classification, the spectral entry associated with the exact $H$-like left mode is identified by its eigenvalue near $-\gamma$ and by its normalized overlap with the projected Hamiltonian vector,
\begin{equation}
	\mathrm{overlap}_H=
	\left|\frac{v_H^\dagger L_\alpha}{\|v_H\|_2\,\|L_\alpha\|_2}\right| .
\end{equation}
We use $\mathrm{overlap}_H>0.985$ as the $H$-like criterion. The steady state is identified by either $|\lambda|<10^{-10}$ or an overlap larger than $0.98$ with the projected steady-state vector. When a numerical visibility coefficient is needed, the relevant left and right modes are first normalized by their Frobenius norms, and $|c_\alpha|<10^{-12}$ is treated as numerical zero.

\section{Supporting data tables}

Table~\ref{tab:fig3-ratios} lists representative largest-size points for Fig.~3 of the main text. For all entries in this table, the thermal reference is governed by the exact \(H\)-like channel, so \(r_{\rm th}=\gamma=0.1\). The selected state decays through the next visible non-\(H\) mode in the \(Q=0\) sector, so \(r_{\rm fast}=\eta r_{\rm th}\). 

\begin{table}[h]
	\caption{Representative slowest-visible-mode rates at the largest sizes shown in Fig.~3 of the main text.}
	\label{tab:fig3-ratios}
	\begin{ruledtabular}
		\begin{tabular}{lccccc}
			model & \(L\) & \(D\) & \(r_{\rm th}\) & \(r_{\rm fast}\) & \(\eta=r_{\rm fast}/r_{\rm th}\)\\
			\hline
			\((2,3)\) model & 15 & 9327 & 0.1000 & 0.1390 & 1.3903\\
			PXP & 18 & 5778 & 0.1000 & 0.1296 & 1.2957\\
			PPXPP & 24 & 9642 & 0.1000 & 0.1901 & 1.9013\\
			PPPXPPP & 28 & 8300 & 0.1000 & 0.1828 & 1.8281\\
			PPPPXPPPP & 30 & 4612 & 0.1000 & 0.1713 & 1.7127\\
		\end{tabular}
	\end{ruledtabular}
\end{table}

Table~\ref{tab:fig4-l18} lists the key slow-mode rates for Fig.~4 of the main text. 

\begin{table}[h]
	\caption{Key rates for slow-mode selection in the PXP model with \(L=18\) and \(\gamma=0.4\).}
	\label{tab:fig4-l18}
	\begin{ruledtabular}
		\begin{tabular}{lccc}
			mode & \(Q\) & rate \(r=-\mathrm{Re}\lambda\) & role\\
			\hline
			steady state & 0 & \(2.6\times10^{-15}\) & stationary\\
			global slowest nonsteady pair & \(\pi\) & 0.2857 & invisible to translationally invariant density matrices\\
			\(H\)-like mode & 0 & 0.4000 & thermal-state slow tail\\
			next \(Q=0\) non-\(H\) mode & 0 & 0.4558 & zero-energy fast-state tail\\
		\end{tabular}
	\end{ruledtabular}
\end{table}

The \(Q=0\) mode-rate ratio inferred from Table~\ref{tab:fig4-l18} is
\begin{equation}
	\frac{r_{\rm next}^{Q=0}}{r_H}
	\simeq
	\frac{0.4558}{0.4000}
	\simeq 1.1395 .
\end{equation}

Table~\ref{tab:selection-classification} summarizes how the initial states used in Fig.~1 of the main text are classified by the two analytic filters. The table is meant as a guide to the logic of the controls: a finite-time crossing alone is not the protected SQME mechanism unless both filters are satisfied and the observed tail is governed by a faster visible \(Q=0\) mode.

\begin{table}[h]
	\caption{Selection-rule classification of representative initial states in the PXP calculations. Here ``yes'' in the \(H\)-filter column means \(\tr(H\rho_0)=0\), so the exact \(H\)-like left mode is invisible.}
	\label{tab:selection-classification}
	\begin{ruledtabular}
		\begin{tabular}{lccc}
			initial state or ensemble & \(T\rho_0T^\dagger=\rho_0\) & \(\tr(H\rho_0)=0\) & interpretation\\
			\hline
			thermal reference \(\rho_{\rm th}\) & yes & no & reference slow tail\\
			selected \(E=0\) scar eigenstate & yes & yes & SQME state\\
			all-zero product state \(\ket{0\cdots0}\) & yes & yes & SQME state\\
			translation-even \(Z_2\) cat state & yes & yes & SQME state\\
			unaveraged density-wave product state & no & yes & control; can see nonzero-\(Q\) sectors\\
			translation-invariant nonzero-energy state & yes & no & control; retains the \(H\)-like channel\\
			generic off-diagonal coherent pair & not guaranteed & not guaranteed & control; no protected asymptotic selection\\
		\end{tabular}
	\end{ruledtabular}
\end{table}

\section{\texorpdfstring{\(Z_q\)}{Zq} cat initial states and preparation route}

The main text uses \(Z_q\) cat states as representative translationally invariant zero-energy initial states. For example, in an even-length periodic chain there are two period-two configurations,
\begin{equation}
	\ket{\mathsf A}=\ket{1010\cdots10},
	\qquad
	\ket{\mathsf B}=\ket{0101\cdots01},
\end{equation}
and the translation-even cat state is
\begin{equation}
	\ket{Z_2^+}=\frac{1}{\sqrt{2}}
	\left(\ket{\mathsf A}+\ket{\mathsf B}\right).
\end{equation}
Its density matrix satisfies \(T\rho T^\dagger=\rho\), equivalently it lies in the \(Q=0\) operator sector.

One symmetric adiabatic preparation route can be implemented with the auxiliary Hamiltonian
\begin{equation}
	\begin{aligned}
		H_{\rm prep}(s)
		&=J_z(s)\sum_j\sigma_j^z\sigma_{j+1}^z
		+J_\perp(s)\sum_j
		\left(\sigma_j^x\sigma_{j+1}^x+\sigma_j^y\sigma_{j+1}^y\right)\\
		&\quad
		-g_x(s)\sum_j\sigma_j^x ,
		\qquad 0\le s\le1,
	\end{aligned}
\end{equation}
where
\begin{equation}
	J_z(s)=J_f\sin^2\frac{\pi s}{2},
	\qquad
	J_\perp(s)=\kappa J_z(s),
\end{equation}
\begin{equation}
	g_x(s)=g_i\cos^2\frac{\pi s}{2}
	+g_f\sin^2\frac{\pi s}{2}.
\end{equation}
We take \(J_f>0\), \(g_i\gg J_f\), \(0<g_f\ll J_f\), and \(0\le\kappa<1\). This path preserves the global \(Z_2\) symmetry
\begin{equation}
	\Pi_x=\prod_j\sigma_j^x,
	\qquad
	[H_{\rm prep}(s),\Pi_x]=0 .
\end{equation}
The initial ground state is \(\ket{+}^{\otimes L}\), which lies in the \(\Pi_x=+1\) sector. At the final point, the antiferromagnetic Ising interaction makes the low-energy subspace mainly spanned by \(\ket{\mathsf A}\) and \(\ket{\mathsf B}\). Adiabatic evolution within the even sector connects the initial state to \(\ket{Z_2^+}\) in a finite system.

The coherent cat state above is the cleanest theoretical representative of a translation-even density-wave state, but it is not required for the selection rule. The rule is formulated for the density matrix. For a period-
\(q\) product pattern \(\ket{\phi_q}\), an experimentally conservative translation-invariant ensemble can be prepared by randomizing the global origin from shot to shot,
\begin{equation}
	\rho_{q,{\rm mix}}
	=\frac{1}{q}\sum_{r=0}^{q-1}
	T^r\ket{\phi_q}\bra{\phi_q}T^{-r},
\end{equation}
where the sum can be replaced by the full translation orbit if the orbit length exceeds \(q\). This mixed state obeys \(T\rho_{q,{\rm mix}}T^\dagger=\rho_{q,{\rm mix}}\). It also obeys \(\tr(H\rho_{q,{\rm mix}})=0\), because \(\rho_{q,{\rm mix}}\) is diagonal in the classical configuration basis whereas the constrained single-flip Hamiltonian has only off-diagonal matrix elements. Thus a phase-coherent cat is useful for producing a pure state, but it is not necessary for the analytic exclusion of the \(H\)-like slow channel. The all-zero state used in Fig.~1 of the main text is even simpler: it is already translation invariant and has zero energy expectation for the same off-diagonal reason.

Table~\ref{tab:zq-states} lists the density-wave patterns used for the cross-model comparisons in Fig.~2 of the main text. In each case, the pattern is allowed by the corresponding constraint and its translation-averaged or cat-state density matrix obeys both \(T\rho_0T^\dagger=\rho_0\) and \(\tr(H\rho_0)=0\).

\begin{table}[h]
	\caption{Representative density-wave states used in the cross-model SQME tests.}
	\label{tab:zq-states}
	\begin{ruledtabular}
		\begin{tabular}{lccc}
			model & constraint & state & representative pattern\\
			\hline
			\((2,3)\) model & no \(111\) strings & \(Z_3\) cat or mixture & \(100100\cdots\)\\
			PPXPP & \(R=2\) blockade & \(Z_3\) cat or mixture & \(100100\cdots\)\\
			PPPXPPP & \(R=3\) blockade & \(Z_4\) cat or mixture & \(10001000\cdots\)\\
			PPPPXPPPP & \(R=4\) blockade & \(Z_5\) cat or mixture & \(1000010000\cdots\)\\
		\end{tabular}
	\end{ruledtabular}
\end{table}

\section{Experimental implementation and observable-level detection}

\subsection{Implementing the constrained dissipative dynamics}

The experimental ingredients required by the theory are a constrained single-spin-flip graph and local dephasing in the Rydberg occupation basis. The coherent part can be generated by resonant Rydberg driving in the hard-blockade regime described above, Eqs.~\eqref{eq:HRy_supp}--\eqref{eq:HR_blockade_supp}. Increasing the blockade radius gives the longer-range $R$-blockade family, while the $(2,3)$ model should be regarded as a distinct constrained graph rather than as PPXPP.
In the idealized master equation used throughout this work, local dephasing is described by jump operators proportional to $n_j$. Equivalently,
\begin{equation}
	L_j=\sqrt{2\gamma}\,n_j,
	\qquad
	\sum_j
	\left(
	L_j\rho L_j^\dagger
	-\frac{1}{2}\{L_j^\dagger L_j,\rho\}
	\right)
	=
	\gamma\sum_j
	\left(
	2n_j\rho n_j-\{n_j,\rho\}
	\right).
	\label{eq:local_dephasing_jump_supp}
\end{equation}
This convention fixes the decay rate of a single-spin-flip coherence and therefore fixes the exact left-mode rate $r_H=\gamma$.
One practical route is to apply controlled site-resolved phase noise,
\begin{equation}
	H_{\rm noise}(t)
	=
	\sum_j \xi_j(t)n_j,
	\qquad
	\langle \xi_i(t)\xi_j(t')\rangle
	=
	2\gamma\delta_{ij}\delta(t-t'),
	\label{eq:phase_noise_supp}
\end{equation}
which produces the same local dephasing term after noise averaging. The locality of the noise is important: purely global detuning noise would dephase only total-Rydberg-number coherences and would not realize the Liouvillian analyzed here.

A closely related experimental route is controlled density measurement through light scattering. In optical-lattice experiments, near-resonant light can induce repeated absorption--spontaneous-emission cycles at a controllable rate. The scattered photons carry local density information and, after tracing out the photon field, suppress coherences between many-body configurations with different site occupations. This mechanism has been used experimentally to study anomalous coherence decay in a dissipative many-body optical-lattice system, where the spontaneous-emission rate was controlled and calibrated by a near-resonant laser~\cite{Exp_new2}. In the corresponding theoretical description, spontaneous light scattering acts as a continuous, strictly local density measurement. Although that experiment was performed in a Bose--Hubbard optical lattice rather than in a Rydberg blockade chain, it provides a direct experimental precedent for engineering number-basis dephasing of the form relevant to Eq.~\eqref{eq:local_dephasing_jump_supp}.

For the Rydberg implementation considered here, the analogous requirement is that the environment resolves the local Rydberg occupation $n_j$ while inducing negligible population relaxation on the timescale of interest. This can be viewed either as local stochastic detuning noise, as in Eq.~\eqref{eq:phase_noise_supp}, or as a weak measurement channel that continuously monitors the Rydberg occupation. In both cases the essential condition is that the dephasing record distinguishes local occupations rather than only the total excitation number. Under this condition, a coherence $|\mu\rangle\langle\nu|$ between two constrained configurations decays at a rate proportional to their Hamming distance, which is precisely the structure used in the proof of $\mathcal L^\dagger(H)=-\gamma H$.

Controlled dissipation in Rydberg platforms has also been demonstrated experimentally. A recent experiment implemented a tunable dissipative channel by resonantly coupling the Rydberg state to a short-lived intermediate state, thereby depumping Rydberg excitations at a controllable rate~\cite{Begoc2025}. The induced dissipation could be varied with the dissipation-laser power and switched on or off during a single experimental cycle. Although this depumping channel is not identical to the pure dephasing dissipator used in our minimal theory, it demonstrates that dissipative timescales in Rydberg experiments can be engineered and calibrated. Together with controlled phase noise and measurement-induced number-basis dephasing, such techniques provide a realistic route to testing Liouvillian mode selection in constrained Rydberg systems.

Finite blockade strength, residual detuning, and long-range interaction tails add diagonal or weak off-graph terms to the ideal constrained Hamiltonian. These terms are experimental imperfections relative to the minimal model, rather than ingredients of the analytic proof. The sharp identity $\mathcal L^\dagger(H)=-\gamma H$ is exact for the projected single-flip Hamiltonian and the local-dephasing dissipator. Experimentally, the relevant test is therefore to calibrate the coherent dynamics inside the blockade subspace, measure the single-flip dephasing rate, and compare the extracted long-time slopes for selected and reference initial ensembles.

\subsection{Initial-state protocol}

The slow-mode selection rule requires two density-matrix conditions,
\begin{equation}
	T\rho_0T^\dagger=\rho_0,
	\qquad
	\tr(H\rho_0)=0.
	\label{eq:experimental_selection_rule_supp}
\end{equation}
The second condition is automatically satisfied by any incoherent mixture of classical product configurations, because $H$ is off diagonal in the computational basis. The first condition can be satisfied either coherently, by preparing a translation-even cat state, or incoherently, by averaging translated density-wave product states over experimental shots. The latter protocol avoids the need to maintain global cat coherence. It prepares the required $Q=0$ density matrix at the ensemble level and is sufficient for testing the Liouvillian visibility mechanism.

A single unaveraged density-wave product state may still show a crossing with the thermal curve at intermediate times. Such a crossing is not the protected SQME mechanism discussed in the main text, because the density matrix contains nonzero-$Q$ components and can couple to additional symmetry sectors. The clean experimental comparison should therefore use either an exactly translation-invariant state or an explicitly shot-averaged translated ensemble when the goal is to test the asymptotic selection rule.
For an experimental test, the two ensembles should be evolved under the same final Liouvillian. The reference ensemble should be finite-temperature-like or energy-biased, with $\tr(H\rho_\beta)\neq0$, so that the exact $H$-like slow channel is visible. The selected ensemble should satisfy Eq.~\eqref{eq:experimental_selection_rule_supp}, for example through a translation-averaged density-wave preparation or the all-zero product state. The essential comparison is therefore not between different dissipators, but between different initial-state visibility conditions under one and the same dissipative dynamics.

\subsection{Observable-level detection}

The trace distance used in the numerical figures is a basis-independent theoretical diagnostic. In few-level quantum systems, this type of distance-to-stationarity can be measured directly by reconstructing the full density matrix. For example, in the trapped-ion observation of the quantum strong Mpemba effect, the time-evolved state $\rho(t)$ was obtained by quantum-state tomography, and the relaxation was characterized by a distance between $\rho(t)$ and the stationary state $\rho_{\rm ss}$~\cite{QME_Exp2}. The strong Mpemba behavior was then identified from the logarithmic decay of this distance: the selected state starts farther from the steady state but approaches it with a faster asymptotic slope.

In many-body Rydberg arrays, however, full many-body tomography and hence a direct measurement of the global trace distance are generally not feasible. The experimentally relevant test should therefore be formulated in terms of accessible observables or basis-dependent statistical distances. This is sufficient for testing the present mechanism, because the predicted speedup is a statement about the visibility of Liouvillian modes: any observable with nonzero response to the leading visible mode will show the corresponding long-time decay rate.

For any measured observable $O$, the deviation from its steady-state value has the left-right Liouvillian expansion
\begin{equation}
	\delta O(t)
	=
	\tr[O(\rho(t)-\rho_{\rm ss})]
	=
	\sum_\alpha c_\alpha o_\alpha e^{\lambda_\alpha t},
	\qquad
	c_\alpha=\tr(L_\alpha^\dagger\Delta\rho_0),
	\qquad
	o_\alpha=\tr(OR_\alpha).
	\label{eq:observable_expansion_supp}
\end{equation}
An observable detects the predicted rate hierarchy if it has nonzero response $o_\alpha$ to the slow mode that is visible to the chosen initial state. Thus the measured signal does not need to reproduce the full trace distance; it only needs to resolve the leading visible decay channel.

A particularly direct diagnostic of the exact slow channel is the energy-like observable $H$ itself. Since $H$ is an exact left eigenoperator,
\begin{equation}
	\frac{d}{dt}\langle H(t)\rangle
	=
	\tr[\mathcal L^\dagger(H)\rho(t)]
	=
	-\gamma \langle H(t)\rangle ,
	\label{eq:H_observable_decay_supp}
\end{equation}
the thermal reference with $\tr(H\rho_\beta)\neq0$ exhibits an $e^{-\gamma t}$ component, whereas selected states satisfying $\tr(H\rho_0)=0$ have no contribution from this channel. In a Rydberg array, $H$ is a sum of constrained single-spin-flip operators and can in principle be accessed through basis-rotated measurements or Ramsey-type probes of local coherences, combined with the local blockade projectors. Such off-diagonal probes are better suited to detecting the exact $H$-like channel than measurements of occupations alone.

Site-resolved fluorescence snapshots nevertheless provide useful many-body diagnostics. Repeated measurements in the Rydberg occupation basis give samples of classical configurations $\mu$, from which one may construct a diagonal, basis-dependent distance to the constrained infinite-temperature distribution,
\begin{equation}
	D_{\rm diag}(t)
	=
	\frac{1}{2}
	\sum_{\mu}
	\left|
	P_t(\mu)-P_{\rm ss}(\mu)
	\right|,
	\qquad
	P_{\rm ss}(\mu)=\frac{1}{D}.
	\label{eq:diagonal_snapshot_distance_supp}
\end{equation}
Here $P_t(\mu)$ is the measured probability distribution over allowed constrained configurations. This quantity is a classical total-variation distance of the measured diagonal ensemble. It is experimentally natural, but it should not be confused with the full quantum trace distance: it is insensitive to coherences that are not mapped onto populations before measurement. Therefore $D_{\rm diag}(t)$ is best viewed as a population-level relaxation diagnostic rather than a direct measurement of the exact $H$-like coherence channel.

Another practical option is an observable-space distance. Let $\{O_a\}$ be a set of measured observables, such as local Rydberg densities, density imbalance, density-wave structure factors, and basis-rotated coherence probes. One can define
\begin{equation}
	D_{\rm obs}(t)
	=
	\left[
	\sum_a w_a
	\left(
	\langle O_a(t)\rangle-\langle O_a\rangle_{\rm ss}
	\right)^2
	\right]^{1/2},
	\label{eq:observable_distance_supp}
\end{equation}
with positive weights $w_a$ chosen according to experimental signal-to-noise. This is not a basis-independent distance between density matrices, but it provides a practical many-body analogue of the distance-to-stationarity measurement used in few-level quantum strong-Mpemba experiments. If the chosen observable set has nonzero response to the leading visible mode, the long-time slope of $\log D_{\rm obs}(t)$ gives the corresponding decay rate.

A more mode-resolved test is possible when approximate slow modes are known theoretically. The trapped-ion experiment reconstructed the density matrix and used the overlaps with individual Liouvillian decay modes to confirm which mode controlled the relaxation~\cite{QME_Exp2}. In a many-body Rydberg setting, complete reconstruction of all mode overlaps is unrealistic, but one can test the same principle selectively: the reference ensemble should display the $H$-like $e^{-\gamma t}$ component in coherence-sensitive observables, whereas the selected ensemble should suppress this component and relax through the next visible $Q=0$ channel.

Suitable diagnostics therefore include density imbalance, density-wave structure factors, local Rydberg densities averaged over symmetry-related sites, diagonal snapshot distances, and, when available, basis-rotated or randomized-measurement probes of coherences. In tweezer-array experiments, site-resolved fluorescence images provide access to local densities, density correlations, and configuration snapshots. In driven-dissipative Rydberg gases, electromagnetically induced transparency has been used to monitor Rydberg population dynamics continuously and nondestructively~\cite{Exp9}. Such measurements do not reconstruct the full density matrix, but they can still reveal the long-time decay rates relevant to the Liouvillian mode-selection mechanism.

Operationally, the experiment should compare the same Liouvillian and the same steady state for two initial ensembles: a finite-temperature-like or energy-biased reference with $\tr(H\rho_\beta)\neq0$, and a translation-invariant zero-energy ensemble with $\tr(H\rho_0)=0$. The thermal reference should exhibit a long-time slope controlled by the exact $H$-like channel with $r_H=\gamma$ in coherence-sensitive probes, while the selected ensemble should exhibit a larger fitted long-time rate $r_{\rm next}^{Q=0}$, provided the measured observable or observable-space distance couples to the next visible mode. The comparison should be made after initial transients and before the finite-size or noise-floor plateau.

\subsection*{Finite-time QME versus SQME}

The distinction between ordinary quantum Mpemba behavior and the strong quantum Mpemba effect is important for interpreting Fig.~1(c) of the main text. A finite-time crossing means that one relaxation curve becomes smaller than another over some time interval. This is a quantum Mpemba effect in the broad sense, but it does not by itself identify the asymptotic slow mode. In this work, SQME means that the candidate fast state removes the slowest visible thermal channel by an analytic selection rule, so that
\begin{equation}
	\frac{\Dtr(\rho_{\rm fast}(t),\rho_{\rm ss})}
	{\Dtr(\rho_{\rm th}(t),\rho_{\rm ss})}
	\sim
	C\,e^{-(r_{\rm fast}-r_{\rm th})t}
	\rightarrow0,
	\qquad r_{\rm fast}>r_{\rm th},
	\label{eq:SQME_asymptotic_supp}
\end{equation}
in the long-time exponential regime. Therefore the control states in Fig.~1(c) may display transient crossings or finite-time acceleration, but they are not counted as SQME unless the density matrix satisfies the translation-invariance condition, the $H$-like overlap condition, and the observed tail is governed by a faster visible $Q=0$ mode. This is why Fig.~1(c) is a control panel, whereas Figs.~1(d)--1(f) are the representative SQME panels.

In practice, both numerical simulations and experiments access only a finite time window. The asymptotic definition above should therefore be translated into an operational finite-time criterion. One should compare the same Liouvillian and the same steady state for two initial ensembles, and identify a time interval after short-time transients but before the numerical plateau or experimental noise floor. In this interval, the selected state should start farther from stationarity but display a larger fitted long-time decay rate than the reference state. Equivalently, for a distance-like diagnostic $\mathcal D_X(t)$, one expects
\begin{equation}
	-\frac{d}{dt}\log \mathcal D_X^{\rm fast}(t)
	>
	-\frac{d}{dt}\log \mathcal D_X^{\rm th}(t)
\end{equation}
over the fitted exponential window. Here $\mathcal D_X$ may be the trace distance in exact numerics, a tomography-based distance in a few-level system, or an observable-level diagnostic in a many-body experiment. The important point is that the fitted slope should reflect the leading visible Liouvillian mode rather than a transient crossing.

This operational viewpoint is consistent with the trapped-ion observation of the quantum strong Mpemba effect, where the density matrix $\rho(t)$ was reconstructed by quantum-state tomography and the relaxation was characterized by a distance between $\rho(t)$ and $\rho_{\rm ss}$~\cite{QME_Exp2}. There, the strong-Mpemba state starts farther from the stationary state than normal initial states, but its logarithmic distance decays with a faster asymptotic slope. The same experiment also reconstructed mode-overlap coefficients and used them to verify the absence of the slowest decaying mode in the selected initial state. These two diagnostics---a faster finite-window decay rate and the suppression of the slowest visible mode---are precisely the experimental counterparts of Eq.~\eqref{eq:SQME_asymptotic_supp}.
For the many-body Rydberg setting considered here, full many-body tomography is generally not required. The same finite-time logic can be applied to coherence-sensitive observables, observable-space distances, or suitable snapshot-based diagnostics. The thermal reference should show a long-time component controlled by the exact $H$-like channel with $r_H=\gamma$, whereas the selected ensemble should suppress this component and relax through the next visible $Q=0$ mode with rate $r_{\rm next}^{Q=0}$. Therefore, the experimental signature of SQME is not merely a crossing of two finite-time curves, but a robust separation of fitted long-time slopes that can be traced back to the analytic invisibility of the $H$-like slow channel.

Our symmetry-protected mechanism should also be distinguished from recently proposed protocol-based acceleration schemes and from the typical strong Mpemba effect emerging in high-dimensional systems. In the present work, the suppression of the slowest decay channel follows from an exact symmetry-enforced selection rule, rather than from dynamical control protocols or concentration-of-measure arguments~\cite{Bao25Reset,Bao26Typicality}.

\section*{ROBUSTNESS TO PREPARATION AND MODEL IMPERFECTIONS}

The exact selection rule assumes that the prepared density matrix satisfies
both filters in Eq.~\eqref{eq:experimental_selection_rule_supp}. However, in
an experiment these conditions may be weakly violated by imperfect state
preparation. A convenient parametrization is
\begin{equation}
	\rho_0^{\rm exp}
	=
	(1-\epsilon)\rho_{\rm sel}
	+
	\epsilon \rho_{\rm err},
	\label{eq:imperfect_state_supp}
\end{equation}
where $\rho_{\rm sel}$ satisfies the two selection conditions and
$\rho_{\rm err}$ denotes an uncontrolled error component. Such an error does
not modify the Liouvillian spectrum, but it can reintroduce a finite visibility
of the modes excluded by the ideal filters. For example, the visibility of the
exact $H$-like left mode becomes
\begin{equation}
	c_H(\rho_0^{\rm exp})
	\propto
	\mathrm{Tr}(H\rho_0^{\rm exp})
	=
	\epsilon\,\mathrm{Tr}(H\rho_{\rm err}),
	\label{eq:imperfect_H_visibility_supp}
\end{equation}
and is therefore suppressed linearly in the preparation error.

Consequently, the imperfectly prepared state contains both the selected fast
tail and a residual slow tail,
\begin{equation}
	\Delta\rho(t)
	\sim
	A_{\rm fast}e^{-r_{\rm fast}t}R_{\rm fast}
	+
	\epsilon A_{\rm slow}e^{-r_{\rm slow}t}R_{\rm slow}
	+\cdots ,
	\label{eq:imperfect_tail_supp}
\end{equation}
where $r_{\rm fast}=r_{\rm next}^{Q=0}$ for the ideal selected state. If the
dominant error couples to the $H$-like mode, then $r_{\rm slow}=\gamma$. If the
error breaks translation symmetry, $r_{\rm slow}$ should instead be taken as
the slowest nonzero-$Q$ mode to which the error component couples. The residual
slow contribution becomes visible only after a crossover time
\begin{equation}
	t_\epsilon^*
	\simeq
	\frac{1}{r_{\rm fast}-r_{\rm slow}}
	\ln
	\left(
	\frac{A_{\rm fast}}{\epsilon A_{\rm slow}}
	\right),
	\qquad
	r_{\rm fast}>r_{\rm slow}.
	\label{eq:imperfect_crossover_supp}
\end{equation}
Thus imperfect preparation does not abruptly destroy the observable
acceleration window. It reintroduces the excluded slow modes with amplitudes
controlled by the preparation error, leading to a logarithmically delayed
crossover at long times. Uniform shot-to-shot averaging over translated
product-state preparations restores translation invariance of the ensemble
density matrix and therefore suppresses the visibility of nonzero-$Q$ slow
modes, without requiring a phase-coherent cat state.

The same logic also applies to weak imperfections of the constrained
Hamiltonian. Let
\begin{equation}
	\mathcal L_\epsilon
	=
	\mathcal L_0+\epsilon\mathcal V,
	\qquad
	\mathcal V(\rho)=-i[V,\rho],
	\label{eq:perturbed_liouvillian_supp}
\end{equation}
where $V$ denotes a small Hamiltonian perturbation. The eigenvalues and
eigenoperators of the Liouvillian are then corrected by standard biorthogonal
perturbation theory. With
\begin{equation}
	\mathcal L_0R_\alpha=\lambda_\alpha R_\alpha,
	\qquad
	\mathcal L_0^\dagger L_\alpha=\lambda_\alpha^*L_\alpha,
	\qquad
	\mathrm{Tr}(L_\alpha^\dagger R_\beta)=\delta_{\alpha\beta},
	\label{eq:biorthogonal_convention_imperfection_supp}
\end{equation}
for an isolated nondegenerate eigenvalue, one obtains
\begin{equation}
	\lambda_\alpha(\epsilon)
	=
	\lambda_\alpha
	+
	\epsilon\mathcal V_{\alpha\alpha}
	+
	\epsilon^2
	\sum_{\beta\neq\alpha}
	\frac{
		\mathcal V_{\alpha\beta}\mathcal V_{\beta\alpha}
	}{
		\lambda_\alpha-\lambda_\beta
	}
	+
	O(\epsilon^3),
	\label{eq:liouvillian_eigenvalue_perturbation_supp}
\end{equation}
where
\begin{equation}
\mathcal V_{\alpha\beta}
=\mathrm{Tr}\!\left[L_\alpha^\dagger \mathcal V(R_\beta)
\right].
	\label{eq:liouvillian_perturbation_matrix_element_supp}
\end{equation}
Thus the rate ratio $\eta=r_{\rm fast}/r_{\rm th}$ is perturbatively stable
as long as the unperturbed spectral margin
$r_{\rm next}^{Q=0}-\gamma$ is larger than the induced rate shifts.

There is, however, an important distinction between different perturbations.
If $V$ preserves the constrained single-spin-flip structure, then
$H_\epsilon=H_0+\epsilon V$ still consists only of single-flip coherences and
therefore remains an exact left eigenmode,
\begin{equation}
	\mathcal L_\epsilon^\dagger(H_\epsilon)
	=
	-\gamma H_\epsilon .
	\label{eq:perturbed_exact_H_mode_supp}
\end{equation}
In this case the energy filter is simply deformed to
$\mathrm{Tr}(H_\epsilon\rho_0)=0$. If the perturbation also preserves
translation symmetry, the $Q=0$ visibility filter remains well defined. By
contrast, diagonal detunings, interaction corrections, or translation-breaking
imperfections generally deform the $H$-like left mode or mix different
operator-momentum sectors, and can reintroduce a slow-mode visibility of order
$\epsilon$. The subsequent long-time dynamics is therefore described by the
same residual-tail form as in Eq.~\eqref{eq:imperfect_tail_supp}, with
$r_{\rm slow}$ now interpreted as the slowest mode made visible by the
Hamiltonian imperfection. The corresponding crossover time is given by
Eq.~\eqref{eq:imperfect_crossover_supp}, up to perturbative shifts of
$r_{\rm fast}$ and $r_{\rm slow}$. Therefore weak Hamiltonian imperfections do
not abruptly destroy the observable acceleration window; they perturbatively
shift the decay rates and reintroduce excluded slow channels with amplitudes
controlled by the imperfection strength.

\end{document}